# Discrete Self-Similarity Of RR Lyrae Stars

# II. Period Spectrum For A Very Large Sample

Robert L. Oldershaw

Earth Sciences Building

Amherst College

Amherst, MA 01002

rloldershaw@amherst.edu

(413) 222-0903

**ABSTRACT**: A recent paper demonstrated a considerable degree of self-similarity between RR Lyrae stars and their atomic scale analogues: excited helium atoms undergoing energy level transitions with $7 \leq n \leq 10$ and $\Delta n = 1$. Discrete self-similarity between these fractal analogues was identified in terms of their masses, radii, oscillation periods, basic morphologies and kinematics. In this second paper on the subject, an extremely large and carefully analyzed sample of RR Lyrae oscillation periods provides further evidence for a unique match between the *predicted* set of discrete periods, based exclusively on the known helium spectrum and the discrete scaling equations of a fractal cosmological paradigm, and the *observed* period spectra of RR Lyrae stars.

I. Introduction

In a previous paper[1] (hereafter Paper I) it was shown that RR Lyrae variable stars appear to share a unique and discrete self-similarity with excited helium atoms undergoing single-level energy transitions between n = 10 and n = 7. These results demonstrated the ability of the Self-Similar Cosmological Paradigm (SSCP), and its discrete self-similar scaling equations, to correlate actual analogue systems from neighboring Scales of nature's discrete fractal hierarchy.

A critical test in Paper I compared an observed RR Lyrae period spectrum with a predicted period spectrum that was based on the known spectrum of helium and the SSCP scaling equations. While the results of this comparison were reasonably compelling, it was necessary to choose a small and highly accurate sample of observed stellar periods for the RR Lyrae test spectrum. In this second paper on the subject of discrete self-similarity between RR Lyrae stars and excited helium atoms, the opposite approach is shown to provide further support for the proposed discrete fractal relationship. An *extremely large*, but carefully analyzed, RR Lyrae period sample is shown to have a spectrum that also agrees very well with the predicted period spectrum.

II. The Predicted Period Spectrum for RR Lyrae Stars

The motivation for expecting a discrete fractal relationship between RR Lyrae stars and excited helium atoms comes from the discrete fractal model of the cosmos mentioned above. A review of the paradigm has been published[2,3] and the author's website[4] provides a comprehensive resource for studying the SSCP. The crucial scaling equations of the SSCP are:

$$R_\Psi = \Lambda R_{\Psi-1} \; , \quad (1)$$

$$T_\Psi = \Lambda T_{\Psi-1} \; \text{and} \quad (2)$$

$$M_\Psi = \Lambda^D M_{\Psi-1} \; , \quad (3)$$

where R, T and M are spatial lengths, temporal "lengths" and masses of analogue systems from neighboring Scales $\Psi$ and $\Psi$-1 of the discrete cosmological hierarchy. The empirically derived dimensionless scaling constants $\Lambda$ and D have values of 5.2 x $10^{17}$ and 3.174, respectively.

Given the fact that the majority of RR Lyrae stars have masses of $\approx$ 0.6 $M_\odot$, Eq. (3) tells us that their unique analogue on the Atomic Scale *must* be helium atoms.[1] The radius range for RR Lyrae stars: 3.7 $R_\odot$ to 7.2 $R_\odot$, can be used to determine[1] that the principal quantum numbers (n) for the helium analogues are almost entirely in the range 7 $\leq$ n $\leq$ 10. The results presented in Paper I also confirmed that the relevant analogue transitions for excited helium atoms are predominantly characterized by systems with angular momentum quantum numbers (l) in the range 0 $\leq$ l $\leq$ 1, and $\Delta$n = 1. With the above information, we can determine[1] the relevant transition periods for helium atoms and convert them into predicted RR Lyrae oscillation periods using Eq. (2). Table 1 presents the data needed for the present test of the proposed discrete self-similarity.

**TABLE 1** Transition Periods for $^4$He [1sns,p → 1s(n-1)s; Singlet and Triplet States; $7 \leq n \leq 10$; $\Delta n = 1$] and Predicted Periods for RR Lyrae Stars

| $n_1 \rightarrow n_2$; $^x$S | $\Delta E$ (atomic units) | Transition Period $1/\nu$ (sec) | Predicted RR Lyrae Oscillation Period (days) |
|---|---|---|---|
| 8p → 7s; $^3$S | 0.00318 | 4.7725 x 10$^{-14}$ | **0.2872** |
| 8p → 7s; $^1$S | 0.00284 | 5.3559 x 10$^{-14}$ | **0.3223** |
| 8s → 7s; $^3$S | 0.00270 | 5.6275 x 10$^{-14}$ | **0.3387** |
| 8s → 7s; $^1$S | 0.00253 | 6.0056 x 10$^{-14}$ | **0.3614** |
| | | | |
| 9p → 8s; $^3$S | 0.00216 | 7.0340 x 10$^{-14}$ | **0.4233** |
| 9p → 8s; $^1$S | 0.00194 | 7.8446 x 10$^{-14}$ | **0.4722** |
| 9s → 8s; $^3$S | 0.00183 | 8.3029 x 10$^{-14}$ | **0.4997** |
| 9s → 8s; $^1$S | 0.00172 | 8.8339 x 10$^{-14}$ | **0.5317** |
| | | | |
| 10p → 9s; $^3$S | 0.00153 | 9.9130 x 10$^{-14}$ | **0.5966** |
| 10p → 9s; $^1$S | 0.00138 | 1.0991 x 10$^{-13}$ | **0.6615** |
| 10s → 9s; $^3$S | 0.00129 | 1.1778 x 10$^{-13}$ | **0.7089** |
| 10s → 9s; $^1$S | 0.00123 | 1.2353 x 10$^{-13}$ | **0.7435** |

It is not claimed that these 12 periods constitute a complete set of predicted periods for RR Lyrae stars since many other closely related transitions are possible for excited helium atoms with $7 \leq n \leq 10$, $0 \leq l \leq 9$ and $\Delta n \leq 3$. However, the 12 periods listed in Table 1 are the dominant periods that specifically apply to the class of RR Lyrae variable stars. This first approximation predicted period spectrum should be sufficient to model the primary features of any sample of RR Lyrae stars, especially when the caveats regarding period shifting and spurious periods are taken into account.[1]

## III. The Empirical Period Spectrum For RR Lyrae Stars

The Optical Gravitational Lensing Experiment (OGLE) has identified a sample of 7612 RR Lyrae variables in the Large Magellanic Cloud (LMC) galaxy.[5] Not only is this sample very large, but also the derived period spectrum for the sample has been prepared with unusual care. The resulting period spectrum, shown in Fig. 1, is a function generated from a "series of ten histograms (bin width 0.01 days) shifted by 0.001 days with respect to each other." The authors further note that "[t]his manner of the period distribution enables precise determination of the most likely periods of all types of the RR Lyrae stars."[5]

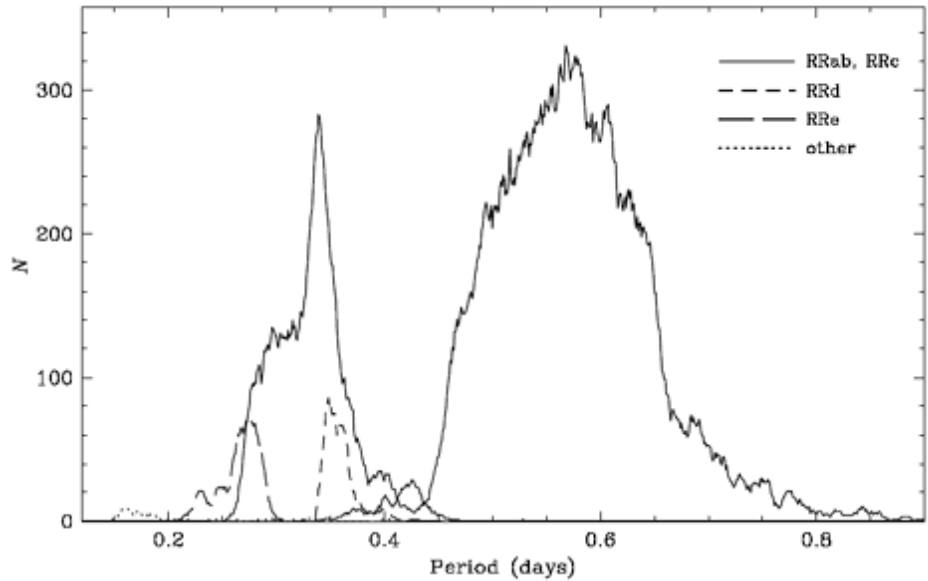

**Figure 1** The observed period spectrum for 7612 RR Lyrae stars in the LMC[5].

Starting from the lowest period end of the spectrum and moving toward the larger period end, the primary features of the OGLE LMC period spectrum of RR Lyrae stars can be identified. These primary features are summarized in Table 2. Below ≈ 0.25 days there are remarkably few RR Lyrae variables. There is a well-defined peak centered at ≈ 0.28 days that is attributed to "second overtone" RRe variables. There is also a definite shoulder between ≈ 0.30 days and ≈ 0.33 days on the main peak of the "first overtone"

RRc variables. The most remarkable feature of the entire spectrum is the large and relatively narrow peak at ≈ 0.34 days. The OGLE team notes[5] that "[t]he most likely period of RRc variables is equal to $<P_c> = 0.339$ days, …". There is a small peak located at about 0.35 days due to RRd variables, whose average period is closer to 0.36 days.

Table 2 Primary Features in the Period Spectrum of the OGLE LMC Sample

| Type of Feature | Location | Predicted Peak(s), Gap | Agreement |
|---|---|---|---|
| Lower Cutoff | ≤ 0.25 days | < 0.28 days | Good |
| Small Peak | 0.28 days | 0.28 days | Excellent |
| Shoulder | 0.30 – 0.33 days | 0.32 days | Good |
| Large Peak | 0.339 days | 0.339 days | Excellent |
| Small Peak | 0.35; <0.36> days | 0.36 days | Good |
| "Blip" | 0.40 days | none | Poor |
| "Blip" | 0.42 days | 0.42 days | Excellent |
| Gap | 0.38 – 0.45 days | > 0.37; < 0.47 days | Good |
| Shoulder | 0.47 days | 0.47 days | Good |
| Broad Peak | 0.46 – 0.67 days | 0.47, 0.50, 0.53, 0.66 days | Fair |
| Upper Cutoff | ~ 0.76 days | 0.74 days | Good |

Another prominent feature of the period spectrum is the pronounced valley in the distribution between ≈ 0.38 days and ≈ 0.45 days. Two tiny "blips" appear at roughly 0.40 days and 0.425 days. At ≈ 0.47 days the main RRab peak rises sharply and remains elevated until sharply decreasing at > 0.65 days. In the case of the smaller sample of RR Lyrae stars for NGC 1835 (see Paper I), the RRab portion of the period spectrum was resolved into individual peaks coincident with the predicted periods. In the case of the huge LMC sample, the resolution of individual RRab peaks is lost. Probably this loss of resolution is due to overlapping among substantial peaks at 0.47 days, 0.50 days, 0.53 days and 0.60 days. Future research efforts should explore various methods for increasing the chances of resolving the RRab peaks in large samples in order to better test for discrete self-similarity in this particular period range. Finally, a broad long-period tail in the distribution stretches until ≈ 0.8 days. As shown in Table 2, the primary features of the observed period spectrum are highly correlated with the discrete spectrum of predicted periods for RR Lyrae stars.

Discrete self-similarity in the period spectra of RR Lyrae stars and excited helium atoms has now been demonstrated for both a small, high-resolution period sample and an extremely large sample of RR Lyrae variable stars. It would appear that the proposed example of discrete cosmological self-similarity is confirmed.